# Effect of Mammalian Tissue Source on the Molecular and Macroscopic Characteristics of UV-Cured Type I Collagen Hydrogel Networks

Charles Brooker [1,2] and Giuseppe Tronci [1,2,*]

[1] Clothworkers' Centre for Textile Materials Innovation for Healthcare (CCTMIH), School of Design, University of Leeds, Leeds LS2 9JT, UK; c.brooker@leeds.ac.uk

[2] School of Dentistry, St. James's University Hospital, University of Leeds, Leeds LS9 7TF, UK

* Correspondence: g.tronci@leeds.ac.uk

## Abstract

The tissue source of type I collagen is critical to ensure scalability and regulation-friendly clinical translation of new medical device prototypes. However, the selection of a commercial source of collagen that fulfils both aforementioned requirements and is compliant with new manufacturing routes is challenging. This study investigates the effect that type I collagen extracted from three different mammalian tissues has on the molecular and macroscopic characteristics of a new UV-cured collagen hydrogel. Pepsin-solubilised bovine atelocollagen (BA) and pepsin-solubilised porcine atelocollagen (PA) were selected as commercially available raw materials associated with varying safety risks and compared with in-house acid-extracted type I collagen from rat tails (CRT). All raw materials displayed the typical dichroic and electrophoretic characteristics of type I collagen, while significantly decreased lysine content was measured on samples of PA. Following covalent functionalisation with 4-vinylbenzyl chloride (4VBC), BA and CRT products generated comparable UV-cured hydrogels with significantly increased averaged gel content ($G \geq 97$ wt.%), while the porcine variants revealed the highest swelling ratio ($SR = 2224 \pm 242$ wt.%) and an order of magnitude reduction in compression modulus ($E_c = 6 \pm 2$ kPa). Collectively, these results support the use of bovine tissues as a chemically viable source of type I collagen for the realisation of UV-cured hydrogels with competitive mechanical properties and covalent network architectures.

1. Introduction

Type I collagen is a major protein component of connective tissues and is largely found in tendon, cartilage, ligament, and skin [1,2]. Molecular alignment, fibrillary organisation, and covalent crosslinking of collagen molecules are key in dictating the macroscopic properties of specific biological tissues in vivo [1] and have been widely pursued for the development of functional collagen-based biomaterials and medical devices. In light of its major role in wound healing and tissue remodelling, collagen has therefore been successfully applied in wound dressings [3,4], guided bone regeneration membranes [5], tendon repair scaffolds [6], pelvic reconstruction meshes [7,8], as well as tracheal [9,10] and corneal [11] implants. Consequently, multiple design strategies, including blending with synthetic polymers [12], fibre spinning [13], and covalent crosslinking [14], have been developed for a range of collagen raw materials generating varying preclinical success. The selection of the tissue source for the extraction of the collagen raw material has increasingly been identified as a key aspect to ensure reproducibility, scalable development, and clinical translation of new medical device prototypes. This is dictated by (i) the animal-induced variation in the chemical composition of collagen; (ii) the recent implementation of the new medical device regulation in the EU [15,16]; and (iii) the increasing efforts of academic institutions worldwide to pursue technology innovation and bridge the gap between the bench and bedside [17].

As a molecule, type I collagen consists of three left-handed polyproline-II chains that are arranged in a right-handed triple helix configuration [18,19]. Each triple helix is a heterotrimeric molecule composed of two alpha-1 polyproline-II chains and one alpha-2 polyproline-II chain. Other than their central regions mediating the formation of the triple helix, each individual alpha chain also contains short nonhelical amino acidic terminations, known as telopeptides, which have been reported to be the sites of the highest interspecies variability in the collagen molecule [20]. Consequently, the immunogenic potential of the collagen molecule is strongly reduced by proteolytic digestion of the telopeptides and covalent crosslinking [21,22]. The latter process is particularly important, aiming to reduce the exposure of the remaining antigenic sites and restore the covalent crosslinks that are broken down during the extraction process [23]. Covalent crosslinking of collagen typically targets the primary amino terminations of lysines, as the most reactive functional group of collagen. From a translational point of view, identifying a collagen raw material with a consistent and relatively high molar content of lysines is critical when aiming to accomplish water-insoluble mechanically compliant materials that are safe in the biological environment and that display retained triple helix conformation.

Type I collagen can be isolated from rat tail tendons; although, the yield of extraction is relatively low and the extracted products are associated with increasing concerns in terms of purity and safety [24]. Other than that, bovine and porcine type I collagens represent readily available collagen raw materials that can be extracted from the mammalian skin and tendon. They are both approved for use in medical devices in the US and Europe, whereby bovine collagen is available in large quantities and accepted by patients who reject porcine products for religious reasons [25]. On the other hand, the use of bovine tissues is associated with the transmission risk of bovine spongiform encephalopathy (BSE), which is a known zoonosis that can lead to severe brain disease in humans [26,27]. These safety risks can be mitigated by extracting collagen from either lower infectivity tissues [26]—e.g., the skin—or by relying on animals from low-BSE-risk countries. Other than that, the safety risks associated with porcine collagen are significantly lower than the ones of the bovine variants, since no cases of transmissible spongiform encephalopathy (TSE) of natural origin have yet been described in pigs and given that pigs present a strong transmission barrier against the propagation of TSEs [28]. Nevertheless, the use of porcine materials and devices is restricted in many countries worldwide, which can potentially emerge as a barrier of clinical uptake.

Other than the aforementioned safety risks, the physiochemical characteristics of collagen raw materials are also of critical importance in the development of new medical devices. Bovine and porcine collagens have been characterised in vitro and revealed differences in terms of swelling, and thermal, and mechanical properties [29,30]. However, these investigations were mainly carried out on in-house extracted, noncommercial collagen raw materials, so their relevance for potential medical device translation is restricted. Furthermore, the manufacturability of the commercial collagen raw materials is also important, aiming to identify regulation-friendly, collagen-based chemical compositions for future medical device development.

This work aims to investigate the molecular characteristics of two commercially available high-purity telopeptide-free collagen raw materials of bovine and porcine origin, respectively, using rat tail collagen as a research-grade comparator. The three abovementioned raw materials were subsequently employed to realise a UV-cured hydrogel network according to a newly established manufacturing route [31]. The molecular and macroscopic properties of the respective hydrogels were quantified and linked to the molecular characteristics of the raw materials, so that controlled structure–property relations could be developed.

## 2. Materials and Methods

### 2.1. Materials

A solution of pepsin-extracted type I BA (6 mg·mL$^{-1}$, 10 mM HCl) was purchased from Collagen Solutions PLC (Glasgow, UK). A phenoxyethanol-supplemented (PE, 0.3 wt.%) aqueous solution of pepsin-extracted type I PA (1 wt.%) was purchased from Koken (Tokyo, Japan). Rat tails were provided post-mortem by the University of Leeds (UK) and employed to extract type I collagen via acidic treatment [32]. Phosphate-buffered saline (PBS, *w/o* Ca$^{2+}$ and Mg$^{2+}$ ions) was purchased from Lonza (Basel, Switzerland). Polysorbate 20 (PS-20), 4-vinylbenzyl chloride (4VBC), and triethylamine (TEA) were purchased from Sigma-Aldrich (Gillingham, UK). 2-Hydroxy-4'-(2-hydroxyethoxy)-2-methylpropiophenone (I2959) was purchased from Fluorochem Limited (Glossop, UK). Absolute ethanol and diethyl ether were purchased from VWR internationals. All other chemicals were purchased from Sigma-Aldrich unless specified.

*2.2. Synthesis of 4VBC-Functionalised Collagen Products*

The solutions of BA (6 mg·mL$^{-1}$, 10 mM HCl) were diluted to a concentration of 3 mg·ml$^{-1}$ via addition of 10 mM HCl and equilibrated to pH 7.4. A total 1 wt.% of PS-20 (with respect to the initial solution volume) and 25 molar excess of both 4VBC and TEA (with respect to the molar content of free amino groups in native collagen) were added. After a 24-h reaction, the mixture was precipitated in a 10-volume excess of pure ethanol and centrifuged (10,000 rpm, 30 min, 4 °C). The reacted collagen pellet was collected and air-dried. To achieve the 4VBC-functionalised PA product, the PE-supplemented PA solution was dialysed against distilled water for 96 h at room temperature using Spectra/Por Standard RC dialysis tubes (MWCO 6-8 kDa) and freeze-dried. The dry product of either PA or CRT was solubilised (0.25 wt.%) in 10 mM HCl and equilibrated to pH 7.4. Following addition of 1 wt.% PS-20, the reaction with 4VBC was carried out as reported above.

*2.3. 2,4,6-trinitrobenzenesulfonic Acid (TNBS) Assay*

TNBS assay ($n = 3$) was used to determine the molar content of lysine amino groups in native and 4VBC-functionalised collagen samples. Briefly, 11 mg of dry sample were mixed with 1 mL of 4 wt.% NaHCO$_3$ and 1 mL of 0.5 vol.% TNBS solution. The mixture was reacted at 40 °C for 4 h, followed by addition of 3 mL of 6 N HCl for one more hour to induce complete sample dissolution. The solution was then cooled down to room temperature, diluted with 5 mL of distilled water, and extracted (×3) with 15 mL diethyl ether to remove any nonreacted TNBS species. An aliquot of 5 mL was collected and diluted in 15 mL of distilled water and the absorbance at 346 nm was recorded using a UV-Vis spectrophotometer (Model 6305, Jenway, Dunmow, UK) against the blank. The molar content of primary amino groups (largely attributed to the side chains of lysine) was calculated via Equation (1):

$$\frac{moles(Lys)}{g(collagen)} = \frac{2 \times Abs(346) \times 0.02}{1.46 \times 10^4 \times b \times x} \tag{1}$$

where *Abs (346 nm)* is the UV absorbance value recorded at 346 nm, 2 is the dilution factor, 0.02 is the volume of the sample solution (in litres), 1.46 × 10$^4$ is the molar absorption coefficient for 2,4,6-trinitrophenyl lysine (in M$^{-1}$·cm$^{-1}$), *b* is the cell path length (1 cm), and *x* is the weight of the dry sample. The degree of 4VBC-mediated collagen functionalisation (*F*) was calculated via Equation (2):

$$F = \left(1 - \frac{moles(Lys)_{4VBC}}{moles(Lys)_{Collagen}}\right) \times 100 \qquad (2)$$

where *mol(Lys)*$_{Collagen}$ and *mol(Lys)*$_{4VBC}$ represent the total molar content of free amino groups in native and 4VBC-functionalised collagen samples, respectively.

*2.4. Circular Dichroism*

Circular dichroism (CD) spectra of native and functionalised samples were acquired with a Chirascan CD spectrometer (Applied Photophysics Ltd., Leatherhead, UK) using 0.2 mg mL$^{-1}$ solutions of collagen in 17.4 mM acetic acid. Sample solutions were collected in quartz cells of 1.0 mm path length, whereby CD spectra were obtained with 2 nm band width and 20 nm min$^{-1}$ scanning speed. The spectrum of the 17.4 mM acetic acid control solution was subtracted from each sample spectrum. Temperature ramp measurements at 221 nm fixed wavelength were conducted from 20 to 75 °C with 20 °C per hour heating rate. The denaturation temperature ($T_d$) was determined as the midpoint of the linearly fitted ($R^2 >$ 0.95) transition region. The mean residue ellipticity ($\theta_{mrw,\lambda}$) was calculated according to Equation (3):

$$\theta_{mrw,\lambda} = \frac{mrw \times \theta_\lambda}{10 \times d \times c} \qquad (3)$$

where *mrw* is the mean residue weight and is equal to 91 g·mol$^{-1}$ for amino acids [33]; $\theta_\lambda$ is the observed ellipticity (degrees) at wavelength *λ*, *d* is the path length (1.0 mm), and *c* is the concentration of collagen (0.2 mg·mL$^{-1}$).

*2.5. Sodium Dodecyl Sulphate-Polyacrylamide Gel Electrophoresis (SDS-PAGE)*

Analytical SDS-PAGE was carried out based upon a previous method [34], except that 20 vol.% glycerol was included in the resolving gel. Proteins were separated using a 12% resolving gel at pH 8.8 and a 4% stacking gel at pH 6.8 using MiniPROTEAN III (Bio-Rad laboratories Ltd., Hertfordshire, UK). Samples of CRT, BA, and PA were dissolved in nonreducing SDS sample buffer (26 mM Tris-HCl pH 6.8, 10% *v/v* glycerol, 0.35% *w/v* SDS, and 0.01% *w/v* bromophenol blue) at 0.1 *w/v*% concentration and heated for 2 min at 90 °C. A total 10 μL of each sample solution were loaded onto 4% stacking gel wells and separated on 12% resolving gels (200 V, 60 min, room temperature). Following electrophoresis, protein bands were visualised after 60 min staining in Coomassie Blue (Avidity Science, Waterford, WI, USA) and 60 min treatment in water. The gels were then imaged using a ChemiDoc Imaging System (Bio-Rad Laboratories, Hercules, CA, USA). The molecular weight of resulting bands was approximated by measuring the relative mobility of the standard protein molecular weight markers.

*2.6. Attenuated Total Reflection Fourier Transform Infrared (ATR–FTIR) Spectroscopy*

ATR–FTIR spectra were recorded on dry samples using a Spectrum BX spotlight spectrometer with a diamond ATR attachment (PerkinElmer, Waltham, MA, USA). Scans were conducted from 4000 to 600 $cm^{-1}$ with 32 repetitions averaged for each spectrum. Resolution and scanning intervals were 4 $cm^{-1}$ and 2 $cm^{-1}$, respectively.

*2.7. Synthesis of UV-Cured Hydrogel Networks*

Hydrogels were prepared by dissolving reacted collagen products in a 10 mM HCl solution supplemented with 1 wt.% I2959. The suspension was cast in a 96-well plate (160 μL per well) or a 12-well plate (1.2 g per well) and UV irradiated at 365 nm (Chromato-Vue C-71, Analytik Jena, Upland, CA, USA) for 15 min on both the top and bottom sides. UV-cured collagen hydrogels were collected with the help of a spatula, thoroughly washed with distilled water, and dehydrated in an increasing series of ethanol-distilled water solutions (0, 10, 20, 40, 60, 80, 3 × 100%). The resulting product was air-dried prior to further use.

*2.8. Swelling Tests*

Dry collagen networks of known mass ($m_d$, $n$ = 9) were individually incubated in aqueous media (10 mM PBS, 2 mL) at 25 °C for 24 h. The swelling ratio (*SR*) was calculated at various time points according to Equation (4):

$$SR = \frac{m_s - m_d}{m_d} \times 100 \tag{4}$$

where $m_s$ is the swollen mass of collagen hydrogel samples.

Other than *SR*, the volumetric swelling (*Q*, *n* = 9) was also calculated according to Equation (5):

$$Q = \left(1 + \frac{\rho_2}{\rho_1}\frac{m_s - m_d}{m_d}\right) \times 100 \tag{5}$$

where $\rho_2$ is the density of the dry collagen network (determined from the values of sample weight and volume), and $\rho_1$ is the density of water at room temperature (~1 g·mL$^{-1}$). Data are presented as mean ± SD.

### 2.9. Gel Content

Gel content was determined as the overall portion of the covalent hydrogel network insoluble in 10 mM HCl solution. Dry, freshly synthesised collagen networks of known mass ($m_d$) were equilibrated in 10 mM HCl solution (2 mL) for 24 h. Resulting hydrogels were air-dried and weighed ($m_1$). The gel content (*G*, *n* = 5) was calculated according to Equation (6):

$$G = \frac{m_1}{m_d} \times 100 \tag{6}$$

Data are presented as mean ± SD.

### 2.10. Compression Tests

PBS-equilibrated hydrogel discs (⌀ = 7 mm, *n* = 8) were compressed at room temperature with a compression rate of 3 mm·min$^{-1}$ (BOSE EnduraTEC ELF 3200, EnduraTEC Systems Corporation, Minnetonka, MN, USA). Stress–compression curves were recorded up to hydrogel failure using a 10 N load cell. The compression modulus was quantified by linear fitting in the strain range associated with 10–15% of maximal stress.

### 2.11. Scanning Electron Microscopy

The cross-sectional morphology of UV-cured collagen networks was inspected via scanning electron microscopy using a Hitachi S-3400N (Hitachi, Tokyo, Japan) after freeze-fracturing in liquid nitrogen and lyophilisation. The SEM was fitted with a tungsten electron source and the secondary electron detector was used. The instrument was operated with an accelerating voltage of 5 kV in a high vacuum with a nominal working distance of 10 mm.

Samples were gold-sputtered using an Agar Auto sputter coater (Agar Scientific, Stansted, UK) prior to examination. Pore diameters ($n$ = 100) were measured using ImageJ software.

*2.12. Statistical Analysis*

Data normality tests were carried out using OriginPro 8.51 software (OriginPro, OriginLab Corporation, Northampton, MA, USA). Statistical differences were determined by one-way ANOVA and the post hoc Tukey test. A $p$ value lower than 0.05 was considered significantly different. Data are presented as mean ± SD.

## 3. Results and Discussion

Collagen samples deriving from three different mammalian tissues (bovine, porcine, and murine) were employed as building blocks for the creation of UV-cured functionalised hydrogels (Figure 1). Telopeptide-free BA and PA, on the one hand, and CRT, on the other hand, were selected as either high-purity collagen or research-grade comparator, respectively. The molecular characteristics of the above raw materials were characterised and linked to hydrogel manufacturability and hydrogel macroscopic properties.

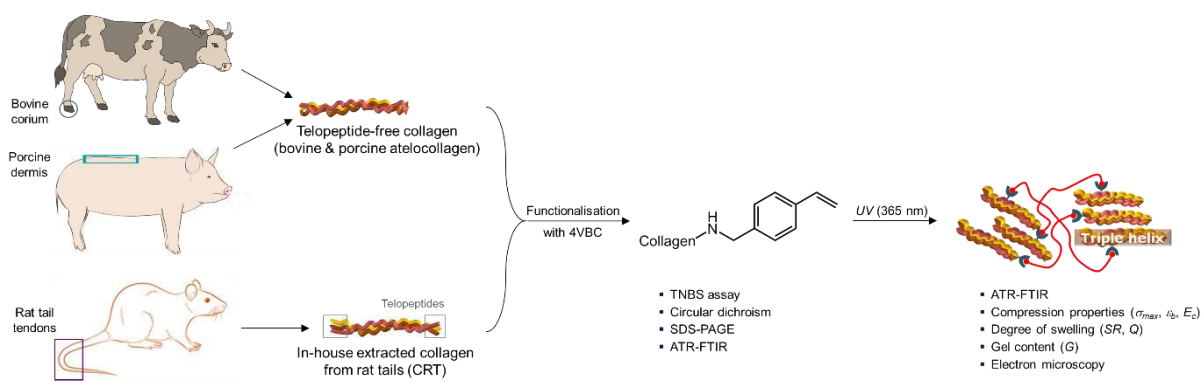

**Figure 1.** Overview of the research strategies and experimental methods. The molecular characteristics of commercially available bovine and porcine atelocollagen were compared with those of research-grade collagen extracted in-house from rat tails. Covalent functionalisation with 4VBC and UV curing were carried out and the effect of raw material molecular characteristics on the manufacturability and macroscopic properties of UV-cured hydrogel networks was assessed.

The reaction of type I collagen with 4VBC was carried out to introduce photoactive residues onto the collagen molecule that could subsequently be photoactivated to generate a UV-cured covalent network. The reaction with 4VBC proceeds via the consumption of primary amino groups, largely attributed to the lysine (Lys) terminations of collagen. A TNBS colorimetric assay was therefore conducted on each mammalian collagen raw material to assess the molar content of Lys, aiming to subsequently quantify $F$ in the 4VBC-reacted collagen product. Samples of both BA and CRT were found to contain an averaged lysine content of at least $2.6 \times 10^{-4}$ moles (Lys) per gram, whereas a lower value was measured in the porcine samples (Table 1), in agreement with previous reports [29]. Following the reaction with 4VBC, all collagen samples displayed a decreased content of lysines, supporting the covalent functionalisation of collagen via the consumption of primary amino groups. Samples of CRT were shown to display the highest average value of $F$ (26 mol.%), whilst insignificantly lower average values were observed in samples of BA ($F$ = 21 mol.%). On the other hand, significant differences were revealed by sample P-4VBC (Table 1), whereby an averaged value of $F$ of just 11 mol.% was obtained, a result that is partially in line with the relatively low lysine content of these porcine samples.

**Table 1.** Lysine content, degree of functionalisation ($F$), ratio of positive to negative magnitude ($RPN$), and denaturation temperature ($T_d$) of native and functionalised collagen samples. CRT—collagen extracted from rat tails, BA—bovine atelocollagen, PA—porcine atelocollagen, R-4VBC—4VBC-functionalised CRT, B-4VBC—4VBC-functionalised BA, P-4VBC—4VBC-functionalised PA, N/A—not applicable.

| Sample ID | $[Lys]/10^{-4} \times$ mol·g$^{-1}$ | $F$/mol.% | $RPN$ | $T_d$/°C |
|---|---|---|---|---|
| CRT | 2.65 ± 0.05 [a] | N/A | 0.12 | 35 |
| BA | 2.60 ± 0.19 | N/A | 0.15 | 34 |
| PA | 2.32 ± 0.04 [a] | N/A | 0.13 | 36 |
| R-4VBC | 1.96 ± 0.19 | 26 ± 7 [a] | 0.12 | 32 |
| B-4VBC | 2.06 ± 0.07 | 21 ± 3 | 0.13 | 35 |
| P-4VBC | 2.06 ± 0.02 | 11 ± 1 [a] | 0.22 | 32 |

[a] Statistically different means ($p < 0.05$).

Together with the lysine content, circular dichroism (CD) spectra were also recorded for both native and functionalised collagen samples in order to elucidate the short-range protein organisation. A typical CD spectrum of type I collagen displays a positive peak at about 221 nm, which is attributed to the presence of right-handed triple helices, and a negative peak at about 197 nm, which relates to the left-handed polyproline-II chains [35]. Both peaks were detected in the CD spectra of all native collagen raw materials (Figure 2A), indicating the nonhydrolysed triple helix organisation of these collagen samples, in contrast to the case of gelatin [32]. Comparable dichroic characteristics were also revealed by the functionalised collagen samples, suggesting the presence of retained 4VBC-functionalised triple helices at this range of *F*. This observation was also confirmed by the values of the magnitude ratio between positive and negative peak intensities (*RPN* = 0.12–0.22, Table 1). The values of *RPN* were comparable regardless of the type of mammalian tissues the respective collagen samples were extracted from or the respective degree of functionalisation. These results indicate minimal variations in triple helix content between both native and 4VBC-functionalised collagens, suggesting a comparable protein organisation in respective UV-cured hydrogels. The fact that there was only a ~13% decrease in *RPN* values following covalent coupling of 4VBC residues (samples of BA) is likely due to the relatively low degree of functionalisation introduced to the collagen (Table 1).

CD measurements were also conducted at varied temperatures and a fixed wavelength ($\lambda$ = 221 nm) to investigate the thermal denaturation of collagen triple helices in both native and functionalised states (Figure 2B). All samples displayed a linear decrease (≥10%) in molar residue ellipticity ($\theta_{mrw,221}$) at about 22–42 °C; thus, the $T_d$ was measured at 32–36 °C (Table 1). Negative values of the CD signal were recorded above ~40 °C, indicating complete triple helix denaturation [32,36].

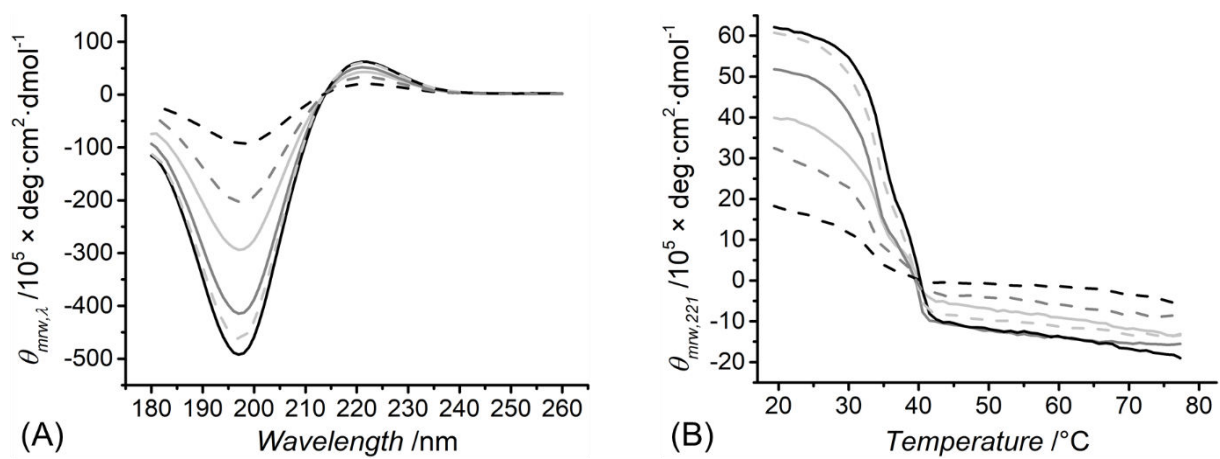

**Figure 2.** (A) Far-UV CD spectra of mean residue ellipticity ($\theta_{mrw,\lambda}$) recorded at constant temperature (20 °C). (B) Temperature ramp CD spectra recorded at a constant wavelength (B, $\lambda$ = 221 nm). Light-grey solid, BA; grey solid, CRT; black solid, PA; light-grey dash, B-4VBC; grey dash, R-4VBC; black dash, P-4VBC. The peaks at 197 and 221 nm are attributed to the presence of left-handed polyproline-II chains and right-handed triple helices, respectively.

These measurements demonstrate the need to introduce covalent crosslinks between collagen molecules aiming to accomplish water-insoluble, collagen-based medical devices in physiological environments.

The chemical composition of the collagen samples was further confirmed by SDS-PAGE [30,37]. The electrophoretic bands associated with monomeric α-chains and cross-linked dimeric β-components were detected at about 124 kDa and 209 kDa, respectively, in line with previous reports [38–40], regardless of the functionalisation and tissue source of collagen (Figure 3). No additional band was observed, indicating that the functionalisation reaction did not cause any detectable degradation of collagen. Since comparable patterns were observed between the sample of CRT and the two commercial samples of BA and PA, this analysis confirmed the validity of the in-house process of acidic collagen extraction.

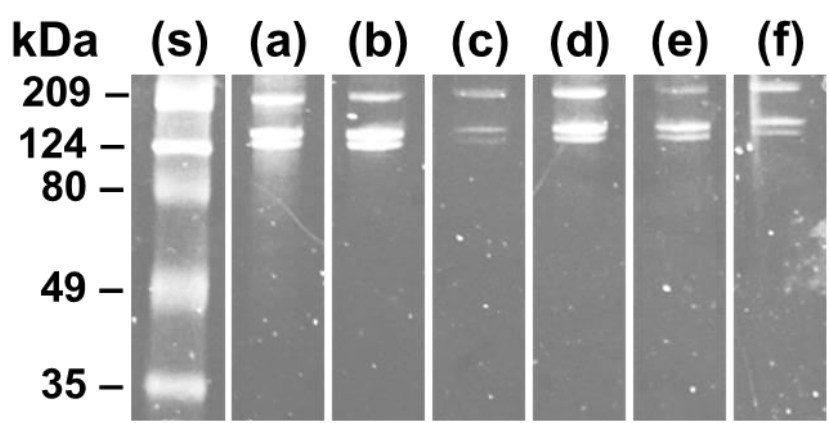

**Figure 3.** SDS-PAGE analysis of the protein markers (s), native collagen samples (a–c), and 4VBC-functionalised collagen samples (d–f). (a) BA; (b) PA; (c) CRT; (d) B-4VBC; (e) P-4VBC; (f) R-4VBC.

Along with CD spectroscopy and SDS-PAGE, ATR–FTIR was employed to further confirm the collagen conformation in the native, functionalised, and UV-cured states. Typically, triple helix collagen is associated with three main amide peaks in the ATR–FTIR spectra [41]. These peaks are the amide I band at ~1650 cm$^{-1}$, the amide II band at ~1550 cm$^{-1}$, and the amide III band centred around 1240 cm$^{-1}$. The amide I band is associated with the peptide C=O stretching, the amide II band relates to both the C-N stretching and N-H bending vibrations, and the amide III band correlates with C-N stretching and N-H bending of amide linkages as well as CH$_2$ wagging vibrations in the glycine backbone and proline side chains. These amide bands are observed in the ATR–FTIR spectra seen in Figure 4. The amide I and II bands, at 1650 cm$^{-1}$ and 1550 cm$^{-1}$, are the most distinct peaks in the spectra of all native and functionalised collagen samples. A less prominent peak is found around 1240 cm$^{-1}$ associated with the amide III band. The fact that these amide peaks are found both before and after functionalisation as well as in the UV-cured state for all three collagen samples suggest that the functionalisation and UV-curing process did not have a significant impact on native protein conformation. Although the presence of vinyl-associated bands at ~1600 cm$^{-1}$ was not detected in the 4VBC-functionalised samples, it is likely that these peaks were masked by the amide I bands related to collagen detected in the same spectral region (~1650 cm$^{-1}$).

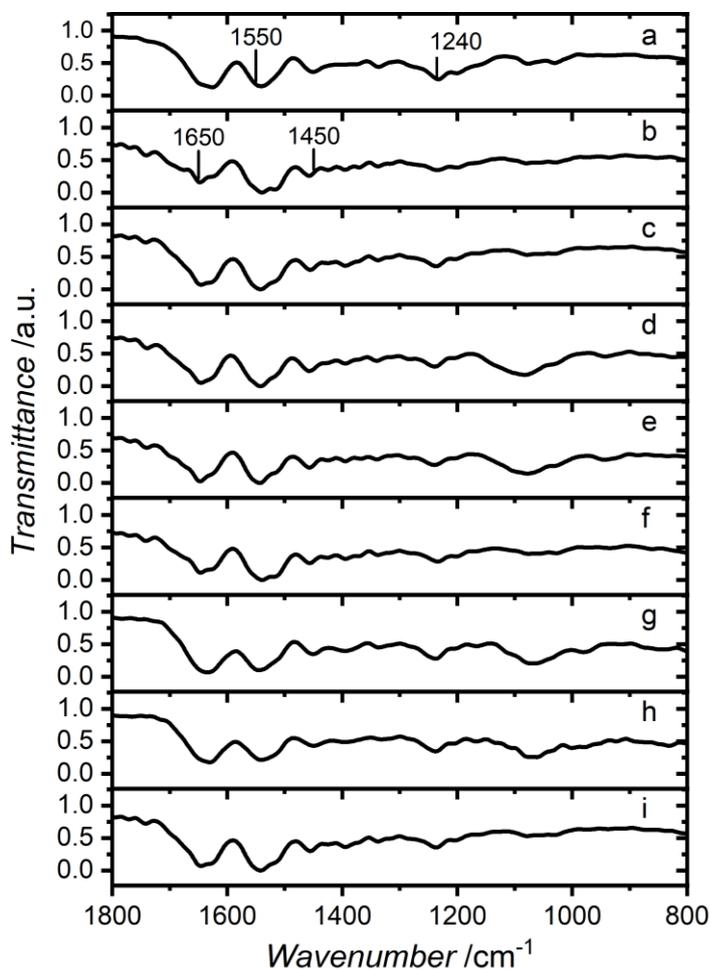

**Figure 4.** ATR–FTIR spectra of native collagen samples (a–c), 4VBC-functionalised collagen samples (d–f), and UV-cured collagen samples (g–i). (a) BA; (b) PA; (c) CRT; (d) B-4VBC; (e) P-4VBC; (f) R-4VBC; (g) UV-cured bovine atelocollagen (B-4VBC*); (h) UV-cured porcine atelocollagen (P-4VBC*); (i) UV-cured rat tail collagen (R-4VBC*). * denotes UV-cured samples.

The mechanical properties of the hydrogels were determined by compression testing. All three UV-cured collagen hydrogels described *J*-shaped stress–compression curves (Figure

5), although the porcine curve appears relatively flat due to the comparatively low stress levels ($\sigma_{max}$ = 4 ± 1 kPa). Hydrogels made of BA (B-4VBC*) and CRT (R-4VBC*) had maximal stresses roughly one order of magnitude greater, with a mean value of 40 and 36 kPa, respectively. The PA hydrogels (P-4VBC*) also had a statistically significant difference in elongation at break, with a mean value of 54%, compared with 37% and 44% for bovine and porcine collagen, respectively. These differences in compressive properties are in agreement with the trends in $F$ and gel content ($G$) recorded in the 4VBC-functionalised collagen products (Table 1) and UV-cured collagen networks (Table 2), respectively.

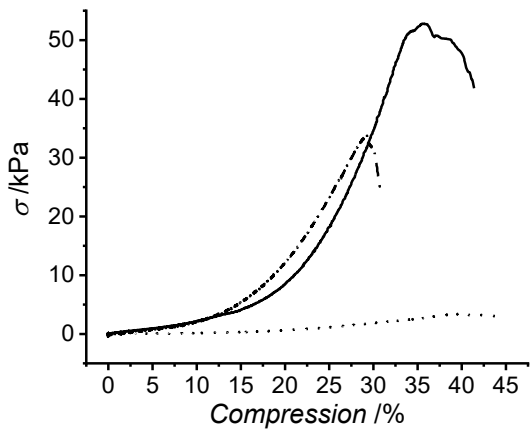

**Figure 5.** Typical stress–compression curves of UV-cured hydrogels: B-4VBC* (black dash dot), R-4VBC* (black solid), and P-4VBC* (black dot).

Indeed, samples of P-4VBC* displayed significantly decreased values of $G$ compared with B-4VBC* and R-4VBC*, suggesting a relatively low crosslink density. On the other hand, the higher values of $F$ observed in samples of B-4VBC and R-4VBC are directly related to the gel content ($G \geq 97$ wt.%) of respective covalent networks and the mechanical properties of the respective hydrogels (Table 2), indicating an increased crosslink density at the molecular scale.

**Table 2.** Macroscopic properties of UV-cured hydrogels made of rat tail collagen (R-4VBC*), bovine atelocollagen (B-4VBC*), and porcine atelocollagen (P-4VBC*). $\sigma_{max}$—maximal stress; $\varepsilon_b$—compression at break; $E_c$—compressive modulus; $SR$—weight-based swelling ratio; $Q$—volumetric swelling; $G$—gel content.

| Sample ID | $\sigma_{max}$ /kPa | $\varepsilon_b$/% | $E_c$/kPa | SR /wt.% | Q /vol.% | G /wt.% |
|---|---|---|---|---|---|---|
| R-4VBC* | 40 ± 14 | 44 ± 7[b] | 67 ± 30 | 1895 ± 292[a] | 1165 ± 179 | 98 ± 1 |
| B-4VBC* | 36 ± 8 | 37 ± 4[c] | 79 ± 25 | 1796 ± 239[b] | 1296 ± 172 | 97 ± 1 |
| P-4VBC* | 4 ± 1[c] | 54 ± 11[b][c] | 6 ± 2[c] | 2224 ± 242[a][b] | 2106 ± 229[c] | 62 ± 5[c] |

[a–c] Statistically different means: [a] $p < 0.05$, [b] $p < 0.01$, [c] $p < 0.001$.

Given the high compatibility of collagen in the aqueous environment, the swelling properties of the UV-cured networks were investigated in PBS. The temporal swelling profiles shown in Figure 6 demonstrate that all three UV-cured 4VBC-functionalised collagen hydrogels showed a remarkable capability of swelling. All types of collagen hydrogel absorbed ten times their weight of media after just 10 min and reached 95% of their equilibrium swelling ratio within 4 h. Hydrogels produced from BA and CRT reached mean equilibrium swelling ratios of ~1800 wt.%, whereas the hydrogels made of PA reached a mean equilibrium swelling ratio of ~2200 wt.% (Table 2). Similar variations were also observed with respect to the volumetric swelling ($Q$), with the porcine hydrogels revealing the highest volumetric expansion ($Q$ = 2106 ± 229 vol.%, Table 2). These differences in swelling properties are expected based on the functionalisation ratios determined via the TNBS assay (Table 1) and given the previously discussed variations in $G$ (Table 2). PA was shown to have a functionalisation ratio almost half that of BA and less than half that of CRT, while respective UV-cured porcine samples revealed a significantly decreased $G$. With this lower degree of crosslinking, there is more flexibility in the collagen molecules, therefore, water diffusion is less hindered [42,43].

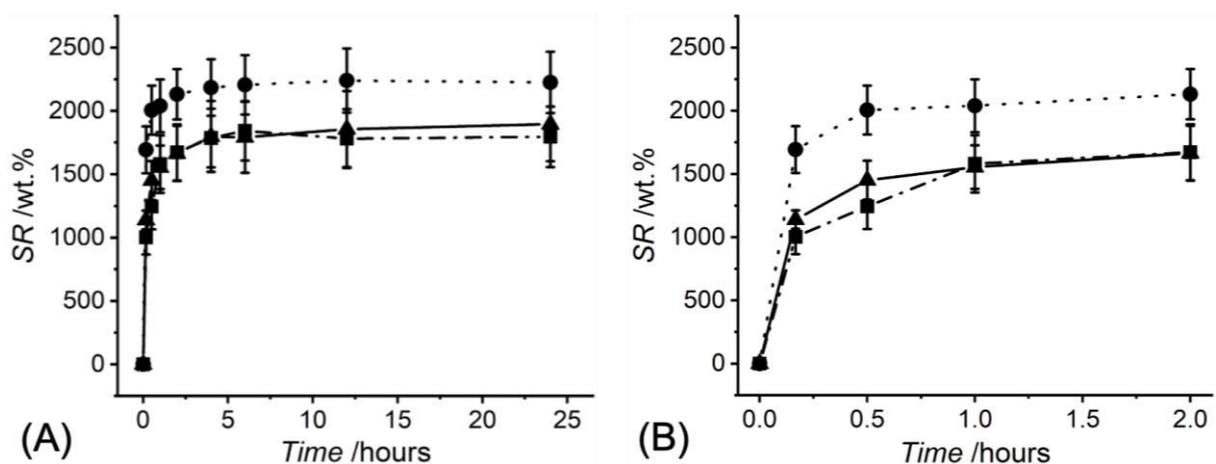

**Figure 6.** Temporal profiles of $SR$ following either 24-h (A) or 2-h (B) incubation of dry UV-cured collagen networks ($n$ = 9) in 10 mM PBS (pH 7.4, 25 °C). Black dot, P-4VBC*; black dash dot, B-4VBC*; black solid, R-4VBC*.

This enables the P-4VBC* samples to take in a larger quantity of media per unit mass, and to take in the media more rapidly, with 90% of the equilibrium swelling ratio absorbed within

30 min, compared with 2 and 4 h for B-4VBC* and R-4VBC* samples, respectively. Consequently, the UV-cured porcine hydrogels display significantly decreased mechanical properties due to the plasticising effect of water.

The cross-sectional morphology of freeze-fractured collagen hydrogels was investigated using SEM to see if there were any morphological features differentiating the collagen hydrogels. Freeze-fractured samples from all three collagen raw materials displayed a series of irregular pores typical of a hydrogel [44], as shown in Figure 7. The topography of the fracture surface varied within samples, with some areas—typically towards the centre of the cross-sectional area—displaying a rough surface made up of fractured fibril and collagen sheets containing pores (Figure 7A,C,E). In other areas, typically closer to the outer hydrogel surface, there is less surface roughness and a greater number of pores (Figure 7B,D,F).

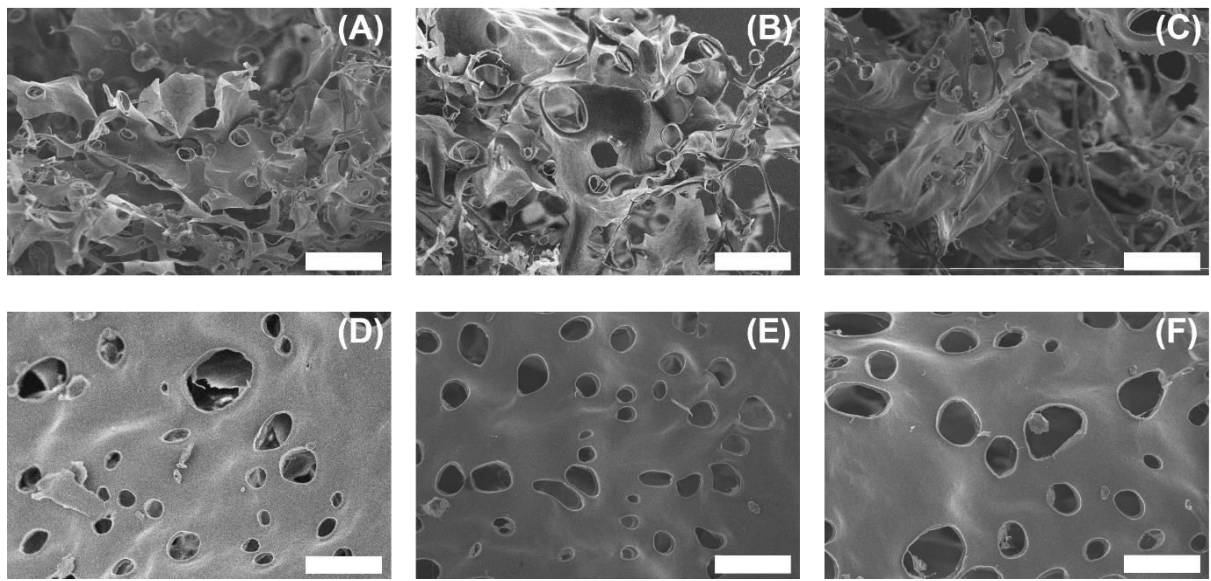

**Figure 7.** SEM Micrographs of freeze-fractured R-4VBC* (A,D); B-4VBC* (B,E); and P-4VBC* hydrogels (C,F), taken from the centre of hydrogel cross-section (A,B,C) and edge of cross-section (D,E,F). All scale bars 100 μm.

This morphological variation within samples could be due to local stress variations within the hydrogel during freeze-fracture. Previously, hydrogels with larger pore sizes, arising from lower degrees of crosslinking, were shown to have an increased swelling ratio [45]. In this study, of the three types of hydrogels, P-4VBC* had the largest mean pore diameter (Ø = 40 ± 16 μm), a statistically significant increase over B-4VBC* (Ø = 28 ± 12 μm) and R-4VBC* (Ø = 34 ± 15 μm), supporting earlier results and the direct effect of pore size on the swelling and compressive properties of hydrogels.

## 4. Conclusions

Type I collagens from three mammalian sources (bovine, porcine, and murine) were covalently functionalised with 4VBC to form UV-cured hydrogels, aiming to develop relationships between the molecular characteristics of the raw material and the macroscopic properties of the resulting hydrogel. While the dichroic, electrophoretic, and spectral characteristics were comparable across the three native collagen samples, lysine content varied depending on the tissue source and was found to be a critical parameter in realising UV-cured hydrogels with controlled macroscopic properties. The lysine content of PA ($[Lys]$ = 2.32 ± 0.04 × $10^{-4}$ mol·$g^{-1}$) was lower than that of both bovine ($[Lys]$ = 2.60 ± 0.19 × $10^{-4}$ mol·$g^{-1}$) and murine ($[Lys]$ = 2.65 ± 0.05 × $10^{-4}$ mol·$g^{-1}$) collagen samples. Consequently, 4VBC-reacted porcine products exhibited the lowest degree of functionalisation ($F$ = 11 ± 1 mol.%) and generated UV-cured hydrogels with reduced gel content ($G$ = 62 ± 5 wt.%) and compression modulus ($E_c$ = 6 ± 2 kPa), as well as increased swelling capabilities ($SR$ = 2224 ± 242 wt.%; $Q$ = 2106 ± 229 vol.%). Other than the porcine variant, bovine and murine hydrogels revealed insignificantly different gel content ($G$ = 97 ± 1– 98 ± 1 wt.%) as well as mechanical ($E_c$ = 67 ± 30–79 ± 25 kPa) and swelling ($SR$ = 1796 ± 239–1895 ± 292 wt.%; $Q$ = 1165 ± 179–1296 ± 172 vol.%) properties. Collectively, these data support the use of bovine collagen as a widely, commercially available and chemically viable raw material for the development of water-insoluble hydrogels with controlled swelling properties and increased mechanical compliance. The comparable molecular and macroscopic characteristics between BA and CRT samples also indicate that in-house extracted murine collagen can be employed as a research-grade proxy of high-purity bovine atelocollagen. Due to their superior macroscopic properties, the abovementioned bovine hydrogels could be appealing for a range of medical devices, including wound healing devices and musculoskeletal tissue scaffolds.

**Author Contributions:** Conceptualization, C.B. and G.T.; Data curation, C.B. and G.T.; Investigation, C.B.; Methodology, C.B.; Project administration, G.T.; Writing – review & editing, C.B. and G.T. All authors have read and agreed to the published version of the manuscript.

**Funding:** This research was funded by the Clothworkers' Company and EPSRC-University of Leeds Impact Acceleration Account.

**Institutional Review Board Statement:** Rat tails were collected post-mortem for the extraction of type I collagen. No animal research was carried out in this study.

**Data Availability Statement:** Not applicable

**Acknowledgments:** The authors gratefully acknowledge Dr. Sarah Myers and Michael Brookes for technical assistance with SDS-PAGE and SEM, respectively. The Clothworkers' Company (London, United Kingdom), the University of Leeds (Leeds, UK) and the Engineering and Physical Sciences Research Council (EPSRC) (Swindon, UK) are also gratefully acknowledged for financial support.

**Conflicts of Interest:** GT is named inventor on a granted patent related to the fabrication of collagen-based materials.